# Photons inside a waveguide as massive particles

**Zhi-Yong Wang[1], Cai-Dong Xiong**

In the paper, we show that there exists a close analogy between the behavior of de Broglie matter waves and that of electromagnetic waves inside a hollow waveguide, such that the guided photons can be treated as free massive particles subject to a relativistic quantum-mechanical equation. Inspired by the effective rest mass of guided photons and the zitterbewegung phenomenon of the Dirac electron, at variance with the well-known Higgs mechanism we present some different heuristic ideas on the origin of mass.

## 1. INTRODUCTION

Traditionally, the theory of waveguides is usually based on classical electromagnetic theory. To study the propagation of electromagnetic wave packets through an undersized waveguide, people have interpreted it as "photon tunneling" via a mathematical analogy between the Helmholtz equation describing evanescent modes and the Schrödinger equation describing a quantum-mechanical tunneling, [(1-3)] where the term related to the cut-off frequency of the waveguide is interpreted as a potential energy.

On the other hand, the similarity between the relativistic dispersion relation of de Broglie matter waves for material particles and that of photons inside a waveguide has been noticed before, [(4-13)] which implies that the term related to the cut-off frequency of the waveguide can be also interpreted as the rest mass of guided photons. However, up to now a systematic description of guided photons based on such an interpretation of cut-off frequency has not been given. In this paper, we will present such a description, by which relativistic quantum mechanics of guided photons is developed.

[1]School of Optoelectronic Information, University of Electronic Science and Technology of China, Chengdu 610054, CHINA; e-mail: zywang@uestc.edu.cn

In the following, the natural units of measurement ($\hbar = c = 1$) is applied, repeated indices must be summed according to the Einstein rule, and the four-dimensional (4D) space-time metric tensor is chosen as $g^{\mu\nu} = \mathrm{diag}(1,-1,-1,-1)$, $\mu,\nu = 0,1,2,3$. For our convenience, let $x^{\mu} = (t, -\boldsymbol{x})$, instead of $x^{\mu} = (t, \boldsymbol{x})$, denote the contravariant position four-vector (and so on), and then in our case $\hat{p}_{\mu} = \mathrm{i}\,\partial/\partial x^{\mu} \equiv \mathrm{i}\partial_{\mu} = \mathrm{i}(\partial_t, -\nabla)$ denote 4D momentum operators.

## 2. SPINOR DESCRIPTION OF THE MAXWELL FIELD

In QED the elementary field quantity is a 4D electromagnetic potential, the quantum theory of the photon is directly a quantum field theory, while the photon's relativistic quantum mechanics (i.e., the first-quantized theory) is absent. However, in spite of QED's great success as well as the traditional conclusion that single photon cannot be localized, [14, 15] there have been many attempts to develop photon wave mechanics which is based on the concept of photon wave function and contains the first-quantized theory of the photon, [16-23] and some recent studies have shown that photons can be localized in space. [24-26] These efforts have both theoretical and practical interests. For our purpose, we will briefly present the first quantized theory of free photons in our formalism.

In vacuum the electric field, $\boldsymbol{E} = (E_1, E_2, E_3)$, and the magnetic field, $\boldsymbol{B} = (B_1, B_2, B_3)$, satisfy the Maxwell equations

$$\nabla \times \boldsymbol{E} = -\partial_t \boldsymbol{B}, \quad \nabla \times \boldsymbol{B} = \partial_t \boldsymbol{E}, \tag{1}$$

$$\nabla \cdot \boldsymbol{E} = 0, \quad \nabla \cdot \boldsymbol{B} = 0. \tag{2}$$

Let $k_\mu = (\omega, \boldsymbol{k})$ denote the 4D momentum of photons, where $\omega$ is the frequency and $\boldsymbol{k}$ the wave-number vector. The momentum-space representation of Eq. (1) can be written as $\boldsymbol{k}\times\boldsymbol{E}(\omega,\boldsymbol{k}) = \omega\boldsymbol{B}(\omega,\boldsymbol{k})$ and $\boldsymbol{k}\times\boldsymbol{B}(\omega,\boldsymbol{k}) = -\omega\boldsymbol{E}(\omega,\boldsymbol{k})$, and then one has $\boldsymbol{k}\cdot\boldsymbol{B}(\omega,\boldsymbol{k}) = 0$ and $\boldsymbol{k}\cdot\boldsymbol{E}(\omega,\boldsymbol{k}) = 0$, which implies that ( $\mathrm{d}^4k \equiv \mathrm{d}\omega\mathrm{d}k_1\mathrm{d}k_2\mathrm{d}k_3$ )

$$\nabla\cdot\boldsymbol{B} = \mathrm{i}\int \boldsymbol{k}\cdot\boldsymbol{B}(\omega,\boldsymbol{k})\exp[-\mathrm{i}(\omega t - \boldsymbol{k}\cdot\boldsymbol{x})]\mathrm{d}^4k = 0\,, \tag{3}$$

$$\nabla\cdot\boldsymbol{E} = \mathrm{i}\int \boldsymbol{k}\cdot\boldsymbol{E}(\omega,\boldsymbol{k})\exp[-\mathrm{i}(\omega t - \boldsymbol{k}\cdot\boldsymbol{x})]\mathrm{d}^4k = 0\,. \tag{4}$$

Therefore, Eq. (2) is actually contained by Eq. (1). The vectors $\boldsymbol{E}$ and $\boldsymbol{B}$ can also be expressed as the column-matrix form: $\boldsymbol{E} = (E_1\ E_2\ E_3)^{\mathrm{T}}$ and $\boldsymbol{B} = (B_1\ B_2\ B_3)^{\mathrm{T}}$ (the superscript T denotes the matrix transpose), by them one can define a 6×1 spinor $\psi(x)$ as follows:

$$\psi = \frac{1}{\sqrt{2}}\begin{pmatrix}\boldsymbol{E}\\ \mathrm{i}\boldsymbol{B}\end{pmatrix}. \tag{5}$$

Moreover, by means of the $3\times 3$ unit matrix $I_{3\times 3}$ and the matrix vector $\boldsymbol{\tau} = (\tau_1, \tau_2, \tau_3)$ with the components

$$\tau_1 = \begin{pmatrix}0 & 0 & 0\\ 0 & 0 & -\mathrm{i}\\ 0 & \mathrm{i} & 0\end{pmatrix},\ \tau_2 = \begin{pmatrix}0 & 0 & \mathrm{i}\\ 0 & 0 & 0\\ -\mathrm{i} & 0 & 0\end{pmatrix},\ \tau_3 = \begin{pmatrix}0 & -\mathrm{i} & 0\\ \mathrm{i} & 0 & 0\\ 0 & 0 & 0\end{pmatrix}, \tag{6}$$

we define the matrix $\beta_\mu = (\beta_0, \boldsymbol{\beta})$ or $\beta^\mu = (\beta^0, -\boldsymbol{\beta})$ ( $\mu = 0,1,2,3$ ), where

$$\beta_0 = \beta^0 \equiv \begin{pmatrix}I_{3\times 3} & 0\\ 0 & -I_{3\times 3}\end{pmatrix},\ \boldsymbol{\beta} = (\beta_1, \beta_2, \beta_3) \equiv \begin{pmatrix}0 & \boldsymbol{\tau}\\ -\boldsymbol{\tau} & 0\end{pmatrix}, \tag{7}$$

Using Eqs. (5)-(7) one can rewrite Eq. (1) as a *Dirac-like equation*

$$\mathrm{i}\beta^\mu\partial_\mu\psi(x) = 0,\ \text{or}\ \ \mathrm{i}\partial_t\psi(x) = \hat{H}\psi(x)\,, \tag{8}$$

where $\hat{H} = -\mathrm{i}\beta_0\boldsymbol{\beta}\cdot\nabla$ represents the Hamiltonian of free photons. Let $\hat{\boldsymbol{L}} = \boldsymbol{x}\times(-\mathrm{i}\nabla)$ be the orbital angular momentum operator, one can easily prove $[\hat{H}, \hat{\boldsymbol{L}} + \boldsymbol{S}] = 0$, where $\boldsymbol{S} = I_{2\times 2} \otimes \boldsymbol{\tau}$ satisfying $\boldsymbol{S}\cdot\boldsymbol{S} = 1(1+1)I_{6\times 6}$ represents the spin matrix of the photon field, where $\otimes$ denotes the direct product. In fact, one can show that the 6×1 spinor $\psi(x)$ transforms according to the $(1,0)\oplus(0,1)$ representation of the Lorentz group (the symbol $\oplus$ represents the direct sum), and can prove that $(\beta^{\mu}\partial_{\mu})(\beta_{\nu}\partial^{\nu}) = \partial^{\mu}\partial_{\mu} + \Omega$, where $\Omega\psi(x) = 0$ is identical with the transverse conditions given by Eq. (2), thus Eq. (8) implies the wave equation $\partial^{\mu}\partial_{\mu}\psi(x) = 0$, where

$$\Omega = I_{2\times 2} \otimes \begin{pmatrix} \nabla_1\nabla_1 & \nabla_1\nabla_2 & \nabla_1\nabla_3 \\ \nabla_2\nabla_1 & \nabla_2\nabla_2 & \nabla_2\nabla_3 \\ \nabla_3\nabla_1 & \nabla_3\nabla_2 & \nabla_3\nabla_3 \end{pmatrix}, \tag{9}$$

where $\nabla_i = \partial/\partial x^i$, $i = 1,2,3$. The Dirac-like equation (8) is valid for all kinds of electromagnetic fields outside a source, and in the first-quantized sense it represents the quantum-mechanical equation of free photons. A more detailed discussion, see Ref. [27]. The Dirac-like equation introduced here is different from those presented in the previous literatures (see, for example, Ref. [28]), in our formalism, the six-component photon wave function should be regarded as a counterpart of the Dirac bispinor (corresponds to the $(1,0)\oplus(0,1)$ representation).

## 3. PHOTONS INSIDE A WAVEGUIDE AS MASSIVE PARTICLES

Though a TEM mode (its electric and magnetic fields are both perpendicular to the direction of propagation) cannot propagate in a single conductor transmission line, a guided wave can be viewed as the superposition of two sets of TEM waves being continually reflected back and forth between perfectly conducting walls and zigzagging down the waveguide, the two sets of TEM waves have the same amplitudes and frequencies, but reverse phases. Usually, the propagation of the electromagnetic wave through an ideal and uniform waveguide is described by a wave equation in (1+1) D space-time. However, in our formalism, we will generally place the waveguide along an arbitrary 3D spatial direction, by which we will show that guided waves have the same behavior as de Broglie matter waves, and in terms of the 6×1 spinor defined by Eq. (5), we obtain a relativistic quantum equation for the guided waves.

In a Cartesian coordinate system spanned by an orthonormal basis $\{\boldsymbol{e}_1, \boldsymbol{e}_2, \boldsymbol{e}_3\}$ with $\boldsymbol{e}_3 = \boldsymbol{e}_1 \times \boldsymbol{e}_2$, we assume that a hollow metallic waveguide is placed along the direction of $\boldsymbol{e}_3$, and the waveguide is a straight rectangular pipe with the transversal dimensions $b_1$ and $b_2$, let $b_1 > b_2$ without loss of generality. It is also assumed that the waveguide is infinitely long and its conductivity is infinite, and the electromagnetic source is localized at infinity. In the Cartesian coordinate system $\{\boldsymbol{e}_1, \boldsymbol{e}_2, \boldsymbol{e}_3\}$, let $k_\mu = (\omega, \boldsymbol{k})$ denote the 4D momentum of photons inside the waveguide, then the wave-number vector is $\boldsymbol{k} = \sum_i \boldsymbol{e}_i k_i = (k_1, k_2, k_3)$ and the frequency $\omega = |\boldsymbol{k}|$, where $k_1 = n\pi/b_1$ and $k_2 = s\pi/b_2$ ($n = 1,2,3...$, $s = 0,1,2...$), and the cutoff frequency of the waveguide is $\omega_{crs} = \sqrt{k_1^2 + k_2^2} = \pi\sqrt{(n/b_1)^2 + (s/b_2)^2}$. For simplicity, we shall restrict our discussion to the lowest-order cutoff frequency $\omega_{\mathrm{c}} = \pi/b_1$. We define the effective rest mass of photons inside the waveguide as $m = \omega_{\mathrm{c}}$, then the photon energy $E = \omega$ satisfies $E^2 = k_3^2 + m^2$. To obtain a Lorentz covariant formulation, let us rechoose a Cartesian coordinate system formed by an orthonormal basis $\{\boldsymbol{a}_1, \boldsymbol{a}_2, \boldsymbol{a}_3\}$ with $\boldsymbol{a}_3 = \boldsymbol{a}_1 \times \boldsymbol{a}_2$, such that in the new coordinate system, one has $\boldsymbol{e}_3 k_3 = \sum_j \boldsymbol{a}_j p_j = \boldsymbol{p}$. That is, in the new coordinate system, the waveguide is put along an arbitrary 3D spatial direction. Let $k_3 \geq 0$ without loss of generality. In the coordinate system $\{\boldsymbol{a}_1, \boldsymbol{a}_2, \boldsymbol{a}_3\}$, one can read $\boldsymbol{k} = \boldsymbol{k}_\perp + \boldsymbol{k}_\parallel$, where (for the lowest-order mode $k_2 = 0$ and $k_1 = \omega_{\mathrm{c}}$)

$$\boldsymbol{k}_{\perp} \equiv \boldsymbol{e}_1 k_1 + \boldsymbol{e}_2 k_2 = \boldsymbol{e}_1 \omega_{\mathrm{c}}\,,\quad \boldsymbol{k}_{\|} = \boldsymbol{p} = \sum\nolimits_i \boldsymbol{a}_i p_i = \boldsymbol{e}_3 k_3\,, \tag{10}$$

stand for $\boldsymbol{k}$ 's components perpendicular and parallel to the waveguide, respectively. Furthermore, one can write the dispersion relation of photons inside the waveguide as $E^2 = \boldsymbol{p}^2 + m^2$, it has the same form as the relativistic dispersion relation of free massive particles, where the cutoff frequency $\omega_{\mathrm{c}} = m = |\boldsymbol{k}_{\perp}|$ plays the role of rest mass, while $\boldsymbol{p}$ represents the momentum of photons along the waveguide, such that the photons moving through the waveguide have an effective 4D momentum $p_{\mathrm{L}\mu} \equiv (E, \boldsymbol{p})$.

According to the waveguide theory, the group velocity ($\boldsymbol{v}_{\mathrm{g}}$) and phase velocity ($\boldsymbol{v}_{\mathrm{p}}$) of photons along the waveguide are, respectively (note that $\hbar = c = 1$)

$$\begin{cases} \boldsymbol{v}_{\mathrm{g}} = \boldsymbol{e}_3 \sqrt{1-(\omega_{\mathrm{c}}/\omega)^2} = \boldsymbol{p}/E \\ \boldsymbol{v}_{\mathrm{p}} = \boldsymbol{e}_3 \Big/ \sqrt{1-(\omega_{\mathrm{c}}/\omega)^2} = \boldsymbol{e}_3\, E/|\boldsymbol{p}| \end{cases}. \tag{11}$$

Then one can obtain the following de Broglie's relations:

$$\begin{cases} \boldsymbol{v}_{\mathrm{g}} \cdot \boldsymbol{v}_{\mathrm{p}} = c^2 = 1 \\ \boldsymbol{p} = \hbar \boldsymbol{k}_{\|} = \boldsymbol{k}_{\|} \\ E = \hbar\omega = \omega = \sqrt{m^2 + \boldsymbol{p}^2} \end{cases}. \tag{12}$$

Using $m = \omega_{\mathrm{c}}$ and Eqs. (11) and (12) one has

$$E = \frac{mc^2}{\sqrt{1-(\boldsymbol{v}_{\mathrm{g}}^2/c^2)}} = \frac{m}{\sqrt{1-\boldsymbol{v}_{\mathrm{g}}^2}}\,. \tag{13}$$

This is exactly the relativistic energy formula. In fact, the group velocity $\boldsymbol{v}_{\mathrm{g}}$ can be viewed as a relative velocity between an observer and a guided photon with the effective rest mass $m = \omega_{\mathrm{c}}$. Eqs. (11)-(13) show that the behavior of guided waves are the same as that of de Broglie matter waves, such that the guided photon can be treated as a free massive particle.

In terms of the effective rest mass $m = \hbar\omega_{\mathrm{c}}/c^2 = \omega_{\mathrm{c}}$, the effective Compton wavelength of guided photons is defined as

$$\lambda\!\!\!^{-}_{c} \equiv \hbar / mc = 1/\omega_{c} \,, \tag{14}$$

As we know, it is impossible to localize a massive particle with a greater precision than its Compton wavelength, which is due to many-particle phenomena. Likewise, inside a hollow waveguide it is impossible to localize a photon along the cross section of the waveguide with a greater precision than its effective Compton wavelength, owing to evanescent-wave phenomena (note that there always exists an inertial reference frame in which the frequency of a propagation mode is equal to the cutoff frequency of the waveguide).

## 4. RELATIVISTIC QUANTUM-MECHANICAL EQUATION OF GUIDED PHOTONS

As we know, a light-like four-vector can be orthogonally decomposed as the sum of a space-like four-vector and a time-like four-vector. In our case, the time-like part of the light-like 4D momentum $k_{\mu} = (\omega, \boldsymbol{k})$ is the effective 4D momentum $p_{\mathrm{L}\mu} = (E, \boldsymbol{p})$, it represents the 4D momentum of photons moving along the waveguide, and is called the traveling-wave or active part of $k_{\mu}$; the space-like part of $k_{\mu}$ is the 4D momentum $p_{\mathrm{T}\mu} \equiv (0, \boldsymbol{k}_{\perp}) = m\eta_{\mu}$ ($\eta_{\mu} \equiv (0, \boldsymbol{k}_{\perp}/m)$ satisfies $\eta_{\mu}\eta^{\mu} = -1$), it contributes to the effective rest mass, and is called the stationary-wave or frozen part of $k_{\mu}$. In other words, as the *rest* energy of photons inside the waveguide (i.e., the energy as the group velocity $\boldsymbol{v}_{\mathrm{g}} = 0$), the effective rest mass arises by freezing out the degrees of freedom related to the transverse motion of photons inside the waveguide. Therefore, we obtain an orthogonal decomposition for $k_{\mu} = (\omega, \boldsymbol{k})$ as follows:

$$k_{\mu} = (\omega, \boldsymbol{k}) = p_{\mathrm{T}\mu} + p_{\mathrm{L}\mu} \,, \quad p_{\mathrm{T}\mu} \equiv (0, \boldsymbol{k}_{\perp}) = m\eta_{\mu} \,, \quad p_{\mathrm{L}\mu} = (E, \boldsymbol{p}) \,. \tag{15}$$

Such an orthogonal decomposition is Lorentz invariant because of $p_{L\mu}p_{T}^{\mu}=0$.

Likewise, as for $x_{\mu}=(t,\boldsymbol{x})$ with $\boldsymbol{x}=\sum_{i}\boldsymbol{e}_{i}x_{i}=(x_1,x_2,x_3)$, in the coordinate system $\{\boldsymbol{a}_1,\boldsymbol{a}_2,\boldsymbol{a}_3\}$ one has $\boldsymbol{e}_3x_3=\sum_{j}\boldsymbol{a}_jr_j=\boldsymbol{r}$, i.e., the 3D vector $\boldsymbol{r}$ is parallel to the waveguide. Let $\boldsymbol{x}_{\perp}\equiv\boldsymbol{e}_1x_1+\boldsymbol{e}_2x_2$, the orthogonal decomposition for $x_{\mu}$ can be written as

$$x_{\mu}=(t,\boldsymbol{x})=x_{T\mu}+x_{L\mu},\ \ x_{T\mu}\equiv(0,\boldsymbol{x}_{\perp}),\ \ x_{L\mu}=(t,\boldsymbol{r}). \tag{16}$$

It is easy to show that

$$k_{\mu}x^{\mu}=(p_{T\mu}+p_{L\mu})(x_{T}^{\mu}+x_{L}^{\mu})=p_{T\mu}x_{T}^{\mu}+p_{L\mu}x_{L}^{\mu}. \tag{17}$$

The operator $\hat{p}_{\mu}=\mathrm{i}\partial_{\mu}=\mathrm{i}\,\partial/\partial x^{\mu}$ represents the totally 4D momentum operator of photons inside the waveguide, while

$$\hat{p}_{L\mu}=\mathrm{i}\partial_{L\mu}=\mathrm{i}\,\partial/\partial x_{L}^{\mu}, \tag{18}$$

represents the 4D momentum operator of photons moving along the waveguide. In spite of the boundary conditions for the waveguide, there are no charges in the free space inside the waveguide, and then where photons should obey the Dirac-like equation $\mathrm{i}\beta^{\mu}\partial_{\mu}\psi(x)=0$. Because of $\psi(x)\sim\exp(-\mathrm{i}k_{\mu}x^{\mu})=\exp[-\mathrm{i}(p_{T\mu}x_{T}^{\mu}+p_{L\mu}x_{L}^{\mu})]$, one has $p_{L\mu}\psi(x)=\mathrm{i}\partial_{L\mu}\psi(x)$, and then

$$\mathrm{i}\beta^{\mu}\partial_{\mu}\psi(x)=\beta^{\mu}k_{\mu}\psi(x)=\beta^{\mu}(p_{L\mu}+p_{T\mu})\psi(x)=\beta^{\mu}(\mathrm{i}\partial_{L\mu}+p_{T\mu})\psi(x). \tag{19}$$

For a given waveguide and a given mode, $k_1=n\pi/b_1$ and $k_2=s\pi/b_2$ ($n=1,2,3...$, $s=0,1,2...$) are fixed, and then $p_{T\mu}=m\eta_{\mu}=(0,k_1,k_2,0)$ is fixed. Using $\mathrm{i}\beta^{\mu}\partial_{\mu}\psi(x)=0$ and $p_{T\mu}=m\eta_{\mu}$, from Eq. (19) one can obtain $\mathrm{i}\beta^{\mu}(\partial_{L\mu}-\mathrm{i}m\eta_{\mu})\psi(x)=0$. Let $\psi(x)=\varphi(x_{L})\exp(-\mathrm{i}p_{T\mu}x_{T}^{\mu})$, obviously $\varphi(x_{L})\sim\exp(-\mathrm{i}p_{L\mu}x_{L}^{\mu})$, and we obtain the Dirac-like equation of photons moving along the waveguide

$$\mathrm{i}\beta^{\mu}(\partial_{L\mu}-\mathrm{i}m\eta_{\mu})\varphi(x_{L})=0. \tag{20}$$

Using Eq. (20) and $\eta^{\mu}\partial_{\mathrm{L}\mu}\varphi(x_{\mathrm{L}})=\partial_{\mathrm{L}\mu}\eta^{\mu}\varphi(x_{\mathrm{L}})=0$ one can obtain the Klein-Gordon equation

$$(\partial_{\mathrm{L}\mu}\partial_{\mathrm{L}}^{\mu}+m^{2})\varphi(x_{\mathrm{L}})=0\,. \tag{21}$$

In the first-quantized sense, Eqs. (20) and (21) serve as the relativistic quantum-mechanical equations of photons moving along the waveguide, while in the second-quantized sense, they are quantum-field-theory equations.

Eqs. (20) and (21) are expressed in the arbitrary coordinate system $\{\boldsymbol{a}_1,\boldsymbol{a}_2,\boldsymbol{a}_3\}$ (associated with a frame wherefrom the waveguide is viewed along an arbitrary 3D spatial direction), they can be simplified in the coordinate system $\{\boldsymbol{e}_1,\boldsymbol{e}_2,\boldsymbol{e}_3\}$ (associated with a frame wherefrom the waveguide is viewed along the $x_3$-axis, and then one has $x_{\mathrm{L}\mu}=(t,0,0,x_3)$ and $m\eta_{\mu}=(0,k_1,k_2,0)$ ). To be specific, in the coordinate system $\{\boldsymbol{e}_1,\boldsymbol{e}_2,\boldsymbol{e}_3\}$, Eq. (20) becomes

$$(\mathrm{i}\beta^{0}\partial_{t}+\mathrm{i}\beta^{3}\partial_{3}+\beta^{1}k_{1}+\beta^{2}k_{2})\varphi(t,x_{3})=0\,. \tag{22}$$

Let $\psi(t,\boldsymbol{x})=\varphi(t,x_3)\exp[-\mathrm{i}(k_1x^1+k_2x^2)]$, one can derive Eq. (8) from Eq. (22), just as one expected. Likewise, let $\varphi(x_{\mathrm{L}})=\varphi(t,x_3)=\exp(\mathrm{i}\omega t)\phi(x_3)$ and $r=x_3$, Eq. (21) assumes the usual form

$$[\partial^{2}/\partial r^{2}+(\omega^{2}-\omega_{\mathrm{c}}^{2})]\phi(r)=0\,. \tag{23}$$

As mentioned before, for the lowest order mode one has $k_1=\omega_{\mathrm{c}}$ and $k_2=0$, and then using $r=x_3$ Eq. (22) can be rewritten as the Schrödinger equation

$$\mathrm{i}\frac{\partial}{\partial t}\varphi(t,r)=(\mathrm{i}\beta_{0}\beta_{3}\frac{\partial}{\partial r}+\beta_{0}\beta_{1}\omega_{\mathrm{c}})\varphi(t,r)\,. \tag{24}$$

To obtain Eq. (24) we have used the fact that in our convention $\beta^{0}=\beta_{0}$ while $\beta^{j}=-\beta_{j}$ ( $j=1,2,3$ ).

## 5. HEURISTIC IDEAS: FROM THE EFFECTIVE REST MASS TO THE ORIGIN OF MASS

The effective rest mass of guided photons inside a hollow waveguide, as the *rest* energy of photons inside the waveguide (i.e., the energy as the group velocity $\boldsymbol{v}_{\mathrm{g}} = 0$), arises by forming standing-waves along the transverse cross-section of the waveguide. In other words, it arises by freezing out the degree of freedom of transverse motion, or, by localizing and confining the electromagnetic energy along the 2D space perpendicular to the waveguide.

In fact, there are other ways of looking at the origin of the effective rest mass of guided photons: Firstly, Eqs. (20) and(21) are covariant under Lorentz boosts along the direction of the waveguide, and the effective rest mass $m = \omega_{\mathrm{c}}$ is an invariant quantity under such Lorentz transformations. In terms of group theory, the effective rest mass of guided photons inside the waveguide originates from the symmetry breaking from SO (1, 3) to SO (1, 1). Secondly, because only those modes in the form of transverse electric (TE) and transverse magnetic (TM) modes can propagate in the waveguide, guided photons inside the waveguide possess the degree of freedom of longitudinal polarization (being parallel to the propagation direction of guided photons along the waveguide), which implies that guided photons have the (effective) rest mass. Finally, guided photons propagating along the length of a hollow waveguide correspond to electromagnetic waves that are reflected back and forth by perfectly conducting walls and zigzag down the waveguide, and their group velocities (also represent their energy velocities, respectively) are the average velocities of velocity-of-light zigzag motion, in such a way they obtain the effective rest mass. Conversely, for a massive particle such as the Dirac electron, its zitterbewegung phenomenon shows that its motion velocity is an average one of velocity-of-light zigzag motion. [(29-35)]

Though the well-known Higgs mechanism has presented a solution to the origin of mass, the Higgs particles predicted theoretically have not been found experimentally. Here we present some heuristic speculations on another potential origin of mass. As mentioned before, there exists a close analogy between the behavior of de Broglie matter waves of massive particles and that of the electromagnetic waves inside a hollow waveguide, which seems to hint that a particle with rest mass corresponds to a standing wave formed by massless fields. Consider that the vacuum medium is described by the vacuum states of quantum fields and then its total momentum vanishes, it is reasonable for us to assume that the vacuum medium as a whole is always resting with respect to all inertial observers. In other words, the relative velocity between the vacuum medium and an arbitrary inertial observer cannot be measured (i.e., it is an unobservable quantity), such that one can think it always vanishes. On the other hand, consider that the velocity of light in vacuum is invariant with respect to all inertial observers, and the eigenvalues of electron's velocity operator are equal to the velocity of light in vacuum, one can present the following hypotheses: the velocity of light in vacuum ( $c=1$ ) and the velocity of the vacuum medium ( $u=0$ ) are only two *genuine* velocities in our universe, they are invariant constants for all inertial frames of reference; all other velocities are the apparent (or average) velocities of massless fields moving in a zigzag manner. Such a zigzag motion, just as the electromagnetic waves that are reflected back and forth by perfectly conducting walls as they propagate along the length of a hollow waveguide, concerns two mutually orthogonal 4D momentum components, i.e., a time-like 4D momentum (called the longitudinal component) and a space-like 4D momentum (called the transverse component), respectively, where the former corresponds to the usual 4D momentum of particles while the latter contributes to the rest mass of particles.

Furthermore, one can guess that a massive particle corresponds to massless fields captured in a bag in the form of standing waves, where the bag consists of the vacuum medium which is unobservable, lossless, and resting with respect to any observer. For an observer moving with a uniform velocity ($\boldsymbol{v}_{\mathrm{g}}$) relative to the standing waves inside the bag, the bag formed by the vacuum medium keeps resting and appears as a resting waveguide extending infinitely along the direction of motion, and the massive particle corresponds to the massless fields moving in a zigzag manner along the waveguide, with the group velocity of $\boldsymbol{v}_{\mathrm{g}}$. The rest mass of the particle depends on the bag's sizes and topological structure, and may also depend on the properties of the vacuum medium inside the bag, as well as the properties of the massless fields captured in the bag (As we know, if a hollow waveguide with the cut-off frequency $\omega_{\mathrm{c}}$ is filled with homogeneous lossless plasma of frequency $\omega_{\mathrm{p}}$, its cut-off frequency becomes $\sqrt{\omega_{\mathrm{c}}^2+\omega_{\mathrm{p}}^2}$, such that photons inside the waveguide have the effective rest mass $m=\sqrt{\omega_{\mathrm{c}}^2+\omega_{\mathrm{p}}^2}$).

## 6. CONCLUSIONS

The behavior of guided waves propagating along a hollow waveguide are closely similar to those of de Broglie matter waves, and the guided photons can be treated as free massive particles, where the effective rest mass of guided photons and the zitterbewegung phenomenon of the Dirac electron together hint that, in addition to the Higgs mechanism, there might also have other origins of mass.

## ACKNOWLEDGMENTS

Project supported by the National Natural Science Foundation of China (Grant No. 60671030).